\documentstyle[preprint,aps,tighten,epsf,floats,float]{revtex}
\newcommand{\be}{\begin{equation}}
\newcommand{\ee}{\end{equation}}
\newcommand{\bea}{\begin{eqnarray}}
\newcommand{\eea}{\end{eqnarray}}
\newcommand{\bPsi}{\bar{\Psi}}
\newcommand{\bh}{\bar{h}}
\newcommand{\bX}{\bar{X}}
\newcommand{\nn}{\nonumber\\}
\newcommand{\Ss}{\scriptsize}

\newcommand{\eqn}[1]{\label{#1}}
\newcommand{\eq}[1]{Eq.~(\ref{#1})}
\newcommand{\eqs}[1]{Eqs.~(\ref{#1})}
\newcommand{\fign}[1]{\label{#1}}
\newcommand{\fig}[1]{Fig.~\ref{#1}}

\begin{document}
\title{\bf Comment on ``Nucleon form factors and a nonpointlike diquark''}
\author{B. Blankleider}
\address{Department of Physics, The Flinders University of South Australia,
Bedford Park, SA 5042, Australia}
\author{A. N. Kvinikhidze\footnote{On leave from Mathematical Institute of
Georgian Academy of Sciences, Tbilisi, Georgia.}}
\address{Department of Physics and Astronomy,
University of Manchester, Manchester, M13 9PL, United Kingdom}
\maketitle
\begin{abstract}
Authors of Phys.\ Rev.\ C {\bf 60}, 062201 (1999) presented a calculation of the
electromagnetic form factors of the nucleon using a diquark ansatz in the
relativistic three-quark Faddeev equations. In this Comment it is pointed out
that the calculations of these form factors stem from a three-quark bound state
current that contains overcounted contributions. The corrected expression for
the three-quark bound state current is derived.
\end{abstract}



\vspace{1cm}

The proper way to include an external photon into a few-body system of strongly
interacting particles described by integral equations has recently been
discussed in detail \cite{nnn1,nnn2}. 
In particular, it has been shown how to avoid the overcounting
problems that tend to plague four-dimensional approaches \cite{nnn1}. The
purpose of this Comment is to point out that just this type of overcounting is
present in the work of Bloch {\em et al.} \cite{bloch} who calculated the
electromagnetic current of the nucleon (and hence form factors), using the
diquark ansatz in a four-dimensional Faddeev integral equation description of a
three-quark system. Moreover, it is shown that the correct expression for the
electromagnetic current consists of just three of the five contributions
calculated in Ref.~\cite{bloch}.

We begin by following Ref.~\cite{nnn2} which is devoted to the discussion of the
electromagnetic current of three identical particles, and is therefore directly
applicable to the present case of a three-quark system. There we used the
gauging of equations method to show that the bound state electromagnetic current
of three identical particles is given by
\be
j^\mu=\bPsi\Gamma^\mu\Psi
\ee
where $\Psi$ ($\bPsi$) is the wave function of the initial (final) three-body
bound state, and $\Gamma^\mu$ is the three-particle electromagnetic vertex
function given by
\be
\Gamma^\mu=\frac{1}{6}\sum_{i=1}^3\left(\Gamma^\mu_iD_{0i}^{-1}
+\frac{1}{2}v_i^\mu d_i^{-1}-\frac{1}{2}v_i\Gamma_i^\mu\right).
\ee
Here $\Gamma^\mu_i$ is the electromagnetic vertex function of the $i$'th
particle, $d_i$ is the propagator of particle $i$, $v_i$ is the two-body
potential between particles $j$ and $k$ ( $ijk$ is a cyclic permutation of 123),
$v^\mu_i$ is the five-point function resulting from the gauging of $v_i$,
and $D_{0i}\equiv d_jd_k$ is the free propagator of particles $j$ and $k$.
Because the bound state wave function $\Psi$ is fully antisymmetric, we can
write
\be
j^\mu=\frac{1}{2}\bPsi\left(\Gamma^\mu_3D_{03}^{-1}
+\frac{1}{2}v_3^\mu d_3^{-1}-\frac{1}{2}v_3\Gamma_3^\mu\right)\Psi. \eqn{j^mu}
\ee
The second term on the right hand side (RHS) of this expression defines
the two-body interaction current contribution
\be
j^\mu_{\mbox{\Ss two-body}}=\frac{1}{4}\bPsi v_3^\mu d_3^{-1}\Psi,
\ee
while the first and third terms together make up the one-body current
contribution to the bound state current. As discussed in Ref.~\cite{nnn1}, the
first term on the RHS of \eq{j^mu} defines an electromagnetic current
\be
j^\mu_{\mbox{\Ss overcount}}=\frac{1}{2}\bPsi\Gamma^\mu_3D_{03}^{-1}\Psi
\ee
which overcounts the one-body current contributions, while the third term
defines a current
\be
j^\mu_{\mbox{\Ss subtract}}=\frac{1}{4}\bPsi v_3\Gamma_3^\mu\Psi
\eqn{subtract}
\ee
which plays the role of a subtraction term in that it removes the overcounted
contributions. Here we shall not be concerned with the two-body interaction
current, but rather, endeavour to examine the cancellations taking place between
the first (``overcount'') and last (``subtract'') terms in detail. Thus we
stress that the correct one-body contribution to the current, also known as
the impulse approximation, is given by
\be
j^\mu_{\mbox{\Ss impulse}}=
j^\mu_{\mbox{\Ss overcount}}-j^\mu_{\mbox{\Ss subtract}}. 
\ee
To reveal these cancellations one writes the bound state wave function in terms
of its Faddeev components
\be
\Psi=\Psi_1+\Psi_2+\Psi_3
\ee
where
\be
\Psi_i=\frac{1}{2}D_{0i}v_i\Psi.  \eqn{Psi_i}
\ee
These components are related through the Faddeev equations
\be
\Psi_i=\frac{1}{2}D_{0i}t_i(\Psi_j+\Psi_k) \eqn{Faddeev}
\ee
where $t_i$ is the $t$ matrix for the $j$-$k$ system,
and for identical fermions obey the symmetry relations \cite{nnn2}
\be
P_{12}\Psi_1=-\Psi_2,\hspace{1cm}P_{13}\Psi_1=-\Psi_3,\hspace{1cm}
P_{23}\Psi_1=-\Psi_1,\hspace{1cm}\mbox{etc.} \eqn{symmetry}
\ee
where $P_{ij}$ is the operator interchanging particles $i$ and $j$. The term
with overcounting is thus
\be
j^\mu_{\mbox{\Ss overcount}}=\frac{1}{2}\left(\bPsi_1+\bPsi_2+\bPsi_3\right)
\Gamma^\mu_3D_{03}^{-1}\left(\Psi_1+\Psi_2+\Psi_3\right)
\ee
which after the use of \eqs{symmetry} becomes a sum of five terms
\bea
j^\mu_{\mbox{\Ss overcount}}&=&\frac{1}{2}\bPsi_3\Gamma^\mu_3D_{03}^{-1}\Psi_3
+\bPsi_3\Gamma^\mu_1D_{01}^{-1}\Psi_3\nn
&+&\bPsi_2\Gamma^\mu_1D_{01}^{-1}\Psi_3
+\bPsi_2\Gamma^\mu_2D_{02}^{-1}\Psi_3+\bPsi_2\Gamma^\mu_3D_{03}^{-1}\Psi_3.
\eqn{five}
\eea
The diquark ansatz used in Ref.~\cite{bloch} is equivalent to invoking the
separable approximation for the two-body $t$ matrix:
\be
t_i=h_i\tau_i\bh_i,
\ee
with $\tau_i$ playing the role of the diquark propagator and $h_i$ describing
the vertex between the diquark and two free quarks. In the case of separable
interactions, it is usual to define the spectator-quasiparticle (quark-diquark)
amplitude $X_i$ through the equation \cite{nnn1}
\be
\Psi_i= G_0 h_i \tau_i X_i
\ee
where $G_0=d_1d_2d_3$.
In terms of these amplitudes the contribution of \eq{five} becomes
\bea
j^\mu_{\mbox{\Ss overcount}}&=&\frac{1}{2}\bX_3d_3\Gamma^\mu_3d_3
\tau_3\left(\bh_3d_1d_2h_3\right)\tau_3 X_3
+\bX_3\tau_3\left(\bh_3d_1\Gamma^\mu_1 d_1 d_2 h_3\right)d_3\tau_3 X_3\nn
&+&\bX_2\tau_2d_2\bh_2d_1\Gamma^\mu_1 d_1h_3d_3\tau_3X_3
+\bX_2\tau_2d_2\bh_2\Gamma^\mu_2d_2d_1h_3d_3\tau_3X_3\nn
&+&\bX_2\tau_2d_2\bh_2d_3\Gamma^\mu_3d_1h_3d_3\tau_3X_3.
\eqn{bloch}
\eea
The five terms summed on the RHS of \eq{bloch} are illustrated in
\fig{fig1}. The last four terms are identical to the contributions
$2\Lambda^i_\mu$ ($i=2,\ldots,5$) of Ref.~\cite{bloch}, while the first term on
the RHS of \eq{bloch} differs from $\Lambda^1_\mu$ only in that our diquark
propagator contains a dressing bubble. With or without this bubble, \eq{bloch}
does not give the correct impulse approximation.
\begin{figure}[H]
\centerline{\epsfxsize=3.7cm\epsfbox{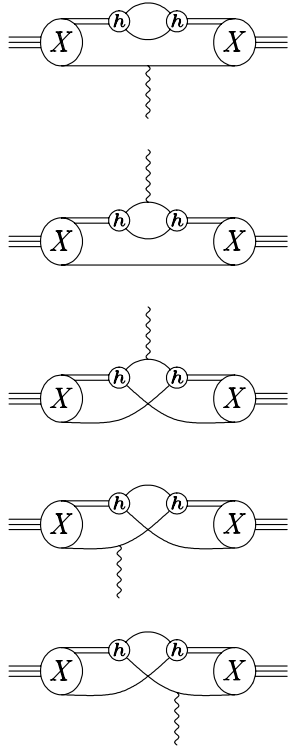}}
\vspace{5mm}
\caption{Illustration of the five terms making up $j^\mu_{\mbox{\Ss overcount}}$
in the case of separable interactions, \eq{bloch}. In the case of a three-quark
system, a single line corresponds to a quark propagator $d_i$, a double line
corresponds to the diquark propagator $\tau_i$, a triple line corresponds to the
three-quark bound state (the nucleon), and the wiggly line indicates the
single-quark electromagnetic current $\Gamma_i^\mu$. The correct impulse
approximation to the three-body bound state current is obtained by removing the
first (top) and fourth (second from the bottom) of these diagrams.}
\fign{fig1}
\end{figure}
With the help of \eq{Psi_i}, the subtraction term of \eq{subtract} 
can be expressed as
\bea
j^\mu_{\mbox{\Ss subtract}}&=&\frac{1}{2}\sum_{i=1}^3
\bPsi_3D_{03}^{-1}\Gamma_3^\mu\Psi_i \nn
&=&\frac{1}{2}\bPsi_3D_{03}^{-1}\Gamma_3^\mu\Psi_3
+\bPsi_2D_{02}^{-1}\Gamma_2^\mu\Psi_3.
\eea
Comparison with \eq{five} shows that the first and fourth terms of \eq{five} are
overcounted.\footnote{Actually the fourth and fifth terms of \eq{five} are
identical, as can easily be shown using \eq{Psieq}. Thus, although we have
singled out the fourth term as the one being overcounted, it should be
understood that overcounting is due to {\em either} the fourth or fifth terms.}
Thus the correct expression for the impulse approximation is
\be
j^\mu_{\mbox{\Ss impulse}}=
\bPsi_3\Gamma^\mu_1D_{01}^{-1}\Psi_3
+\bPsi_2\Gamma^\mu_1D_{01}^{-1}\Psi_3
+\bPsi_2\Gamma^\mu_3D_{03}^{-1}\Psi_3. \eqn{impulse}
\ee
For the work of Ref.~\cite{bloch}, this means that the correct impulse
approximation is given by the sum of their $\Lambda^2_\mu$, $\Lambda^3_\mu$, 
and $\Lambda^5_\mu$ only, and not, as claimed in their work, by the sum of all
five $\Lambda^i_\mu$'s. Diagrammatically this means that the correct impulse
approximation to the nucleon current in the diquark model corresponds to the
sum of the second, third, and fifth diagrams of \fig{fig1}. 

A further comment regarding Ref.~\cite{bloch} concerns the numerical values
obtained for the contributions $\Lambda^1_\mu$ and $\Lambda^5_\mu$.
By using the symmetry properties of \eqs{symmetry}, one can rewrite
the Faddeev equations, \eqs{Faddeev}, as
\be
\Psi_i=D_{0i}t_i\Psi_j  \eqn{Psieq}
\ee
where $i\ne j$. For separable interactions this implies that the amplitudes
$X_i$ satisfy the equations
\be
X_i=\bh_i D_{0i}h_j\tau_j X_j     \eqn{Xeq}
\ee
where $i\ne j$. Using the time-reversed version of these equations one obtains
$\bX_3=\bX_2\tau_2\bh_2 d_1 d_2 h_3$ which can be used to simplify
the last term of \eq{bloch}:
\be
\bX_2\tau_2d_2\bh_2d_3\Gamma^\mu_3d_1h_3d_3\tau_3X_3
=\bX_3 d_3\Gamma^\mu_3 d_3\tau_3X_3.
\ee
The RHS of this equation is just $2\Lambda^1_\mu$ of Ref.~\cite{bloch} and we
have therefore shown that
\be
\Lambda^1_\mu=\Lambda^5_\mu.
\ee
This equality appears not to be reflected in the numerical results
of Ref.~\cite{bloch} as is evident from their Table II.

Finally, we note that the errors of Ref.~\cite{bloch} have been perpetuated in a
recent preprint \cite{bloch2}.

\end{document}